\documentclass[pre,preprint,preprintnumbers,amsmath,amssymb]{revtex4}
\usepackage{amsmath,graphics,epsfig,epstopdf,fancyhdr,dcolumn,bm}

\newcommand{\ad} {\alpha_d}

\newcommand{\kld}{k\lambda_D}

\newcommand{\R}{\text{Re}(\chi)}
\newcommand{\Rf}{\chi^r_{\text{eff}}}
\newcommand{\dRv}{\partial_\omega\chi^r_{\text{env}}}
\newcommand{\Rv}{\chi^r_{\text{env}}}
\newcommand{\I}{\text{Im}(\chi)}

\newcommand{\Vl}{V_l}
\newcommand {\sina} {\langle \sin(\varphi) \rangle}
\newcommand{\wpe}{\omega_{pe}}
\newcommand{\Ic}{\text{Im}(\chi_a)}

\newcommand{\ksl}{k_s^{lin}}
\newcommand{\kll}{k_l^{lin}}

\newcommand{\wsl}{\omega_s^{lin}}
\newcommand{\wll}{\omega_l^{lin}}

\begin {document}
\title{Nonlinear envelope equation and nonlinear Landau damping rate for a driven electron plasma wave} \author{Didier B\'enisti$^1$}
\email{didier.benisti@cea.fr} \author{Olivier Morice$^1$}
\author{Laurent Gremillet$^1$} 
\author {David J. Strozzi$^2$}\affiliation{ $^{1}$CEA, DAM, DIF F-91297 Arpajon, France. \\
  $^{2}$Lawrence Livermore National Laboratory, University of
  California, Livermore, CA 94550} \date{\today}
\begin{abstract}
In this paper, we provide a theoretical description, and calculate, the nonlinear frequency shift, group velocity and collionless damping rate, $\nu$, of a driven electron plasma wave (EPW). All these quantities, whose physical content will be discussed, are identified as terms of an envelope equation allowing one to predict how efficiently an EPW may be externally driven. This envelope equation is derived directly from Gauss law and from the investigation of the nonlinear electron motion, provided that the time and space rates of variation of the EPW amplitude, $E_p$, are small compared to the plasma frequency or the inverse of the Debye length. $\nu$ arises within the EPW envelope equation as more complicated an operator than a plain damping rate, and may only be viewed as such because $(\nu E_p)/E_p$ remains nearly constant before abruptly dropping to zero. We provide a practical analytic formula for $\nu$ and show, without resorting to complex contour deformation, that in the limit $E_p \rightarrow 0$, $\nu$ is nothing but the Landau damping rate. We then term $\nu$ the ``nonlinear Landau damping rate'' of the driven plasma wave. As for the nonlinear frequency shift of the EPW, it is also derived theoretically and found to assume values significantly different from previously published ones, assuming that the wave is freely propagating. Moreover, we find no limitation in $k \lambda_D$, $k$ being the plasma wavenumber and $\lambda_D$ the Debye length, for a solution to the dispertion relation to exist, and want to stress here the importance of specifying how an EPW is generated to discuss its properties. Our theoretical predictions are in excellent agreement with results inferred from Vlasov simulations of stimulated Raman scattering (SRS), and an application of our theory to the study of SRS is presented. 
\end{abstract}
 \maketitle
\section{Introduction}
Landau damping is a linear, collisionless process, resulting from the global acceleration of electrons by an electrostatic wave. Indeed, in the linear regime, an electron plasma wave (EPW) with phase velocity $v_\phi$ globally  accelerates the electrons of initial velocity $v_0<v_\phi$, and decelerates the other ones. When this leads to an overall acceleration of the electrons by the wave, as for example in an initially Maxwellian plasma then, because of energy (or momentum) conservation, the plasma wave damps while, in opposite regime when the electrons are globally decelerated,  the wave grows unstable. The damping, or growth rate, $\nu_L$, of the EPW in the linear regime was first derived by Landau in his famous 1946 paper Ref. \cite{landau}. While addressing the growth of the EPW was rather straightforward, Landau had to use complex contour deformation and analytic continuation to derive the damping rate, a technique which initially shed some doubts into plasma physicists minds as regards the validity of Landau's calculation, all the more as Landau never clearly discussed the physics of the damping. Landau damping, or growth, is  predominantly due to the nearly resonant electrons, those whose initial velocity $v_0$ is such that $\vert v_0-v_\phi \vert \alt \nu_L/k$, where $k$ is the plasma wave number (while the exactly resonant ones, such that $v_0=v_\phi$, do not contribute to it). Then, as is well known, if $\nu_L \ll \omega_{pe}$, where $\omega_{pe}$ is the plasma frequency, $\nu_L$ is approximately proportional to the derivative, $f'_0(v_\phi)$, of the electron distribution function in the limit of a vanishing field amplitude.

A nonlinear counterpart of $\nu_L$ was first calculated by O'Neil in Ref. \cite{oneil}, who considered an electron plasma wave of constant and uniform amplitude, $E_0$, which grew infinitely quickly in an initially Maxwellian plasma. When $\omega_B \gg \nu_L$, where $\omega_B=\sqrt{ekE_0/m}$, $-e$ being the electron charge and $m$ its mass, most of the nearly resonant electrons are trapped and oscillate in the wave trough. Within one oscillation period, a trapped electron neither gains nor loses energy in the wave frame, so that the mechanism which gave rise to Landau damping vanishes, and so does the damping rate after a few oscillations at a frequency close to $\omega_B$, as shown by O'Neil. 

A countless number of papers, addressing both the linear and nonlinear regimes, have been written since these two seminal works were published. In the linear regime, the physics of Landau damping was extensively discussed (see Ref. \cite{nicholson,elskens} and references therein), and new derivations of Landau damping which did not resort to complex contour deformation, or which extended Landau's result to non smooth initial distribution functions (as is the case for a real plasma made of discrete particles) were found (see Ref. \cite{elskens}). Moreover, very recently, Belmont \textit{et al.} showed in Ref.  \cite{belmont} the very unexpected result that an EPW could damp at a rate different, and lower, than that derived by Landau, provided that this wave was excited from noise in such a way that the electron distribution function had a complex pole in velocity space. This shows the importance of specifying the way an EPW has been created in order to correctly discuss its physics properties and, in particular, to correctly calculate its complex frequency. In this paper, we provide a derivation of the Landau damping rate which does not resort on complex contour deformation and which, we believe, is quite simple. This moreover allows us to discuss the ability to excite a plasma wave in such a way that it decays at a non-Landau rate.

In the nonlinear regime, several papers recently discussed the very work of O'Neil, eventually leading to its experimental check (see Ref. \cite{danielson} and references therein). Although the situation considered by O'Neil is physical and could be reproduced experimentally, it is not the most general one since a plasma wave amplitude usually depends on both space and time, even when this wave induces nonlinear electron motion. Generalizing O'Neil's results has been a long standing problem in plasma physics, which we address in this paper. In particular, we provide an analytic expression, supported by numerical results, for the nonlinear collisionless damping rate, $\nu$, of a plasma wave whose amplitude may vary in space and time, in the limit of non relativistic electron motion and slow amplitude variations. We moreover restrict to a driven plasma wave for the following reasons. First, only if an EPW is externally driven may it grow in an initially Maxwellian plasma and may global electron acceleration, at the origin of Landau damping, occur. Second, for a driven wave, the initial conditions can be defined unambiguously and, in particular, one may assume that the plasma wave amplitude is initially at a noise level. This allows one to discuss the generality of previous results, regarding the nonlinear dispersion relation of an EPW, derived by assuming that the was was freely propagating. Third, our work directly applies to stimulated Raman scattering (SRS), which is studied as a tool for amplification of electromagnetic radiation, but which may also be detrimental for an inertial confinement device such as the Laser M\'egaJoule \cite{cavailler}, because it may induce the reflection of a substantial part of the incident laser energy.  Now, recent numerical \cite{vu,strozzi} and experimental \cite{montgomerry} papers on SRS reported reflectivities far above what could be inferred from linear theory. This so-called ``kinetic inflation'' was attributed to the nonlinear reduction of the Landau damping rate, although no theory, nor analytic formula, was available to support this assumption. The present paper addresses this issue and discusses in detail the derivation and physics of the very recent results, published in Ref. \cite{benisti09}. 

There are several caveats in trying to define, and calculate, the nonlinear collisionless damping rate, $\nu$, of a driven wave. For example, one cannot that easily use energy conservation as O'Neil did, nor even momentum conservation, to derive $\nu$, because the electrons are accelerated by both, the drive and the plasma wave. It is usually argued that the plasma wave amplitude, $E_p$, is much larger than that, $E_d$, of the laser drive, and this argument has been used by Yampolsky and Fisch in Ref. \cite{yampolsky} to derive a set of equations from which $\nu$ could be derived numerically, in case of a purely time growing wave. The relative values of $E_p$ and $E_d$ has actually been investigated in detail in Ref. \cite{benisti08} where it has been shown that only in the nonlinear regime when $\nu \approx 0$, or in the linear regime when the Landau damping rate is small enough, is $E_p \gg E_d$. Moreover, even in these regimes, only the space integrated energy, or momentum, is conserved, and these global quantities are not easily related to $\nu$ which is defined locally. Since $\nu$ is not easily calculated using conservation laws, in this paper, we will derive it from Gauss law, which is unambiguous. Using the electron susceptibility, $\chi$, introduced in Ref. \cite{benisti07}, and whose definition will be recalled in Section \ref{exp}, we find that, provided that  when $\R \approx -1$ and $\vert \I \vert \ll 1$ (which are easily achieved conditions), $E_p$ is related to $E_d$ and to the dephasing $\delta \varphi$ between the plasma wave and the external drive by the equation,
\begin{equation}
\label{eq1}
\I E_p-k^{-1} \partial_xE_p = E_d \cos(\delta \varphi).
\end{equation}
Eq. (\ref{eq1}) tells us how efficiently an electron plasma wave may be driven, which is an important issue since our work was primarily motivated by the estimating of Raman reflectivity in fusion devices. To this respect, the nonlinear derivation of $\I$, which will be discussed in detail throughout this paper, is essential since it is clear, from Eq. (\ref{eq1}), that a nonlinear decrease of $\I$ would enhance the driving of the EPW and, hence, SRS. Now, it is also clear that, while it is driven, an EPW accelerates the plasma electrons exactly the same way as if it were freely propagating, which hampers its growth. The effectiveness of the EPW drive therefore significantly depends on the rate of energy, or momentum, transfer from the wave to the electrons, a process akin to that giving rise to the Landau damping of a freely propagating wave. We would like to make this more transparent by writing Eq. (\ref{eq1}) in terms of an envelope equation of the form,
\begin{equation}
\label{eq2}
\partial_t E_p+v_g \partial_x E_p+\nu E_p=E_d \cos (\delta \varphi)/\dRv.
\end{equation}
Then, $v_g$ would be called the group velocity of the plasma wave, and $\nu$ its nonlinear Landau damping rate. In this paper, we indeed show how to derive Eq. (\ref{eq2}) from Eq. (\ref{eq1}) and we actually provide an analytic formula for $\nu$, that matches the Landau damping rate, $\nu_L$, in the limit of vanishing field amplitudes. We moreover show that $\nu$, which depends on both the wave amplitude and its space and time variations, may be viewed as a plain damping rate because it assumes nearly constant values before abruptly dropping to zero. Then, not only is Eq. (\ref{eq2}) physically more transparent than Eq. (\ref{eq1}) but it is also easier to solve numerically to get, for example, quantitative estimates for Raman reflectivity. It is however important to note that the physical meanings of $\nu$ and $v_g$ are not as obvious as for a freely propagating wave. Indeed, usually, the maximum of a driven plasma wave packet does not travel at $v_g$. Moreover, the amplitude of the driven EPW does not decrease at rate $\nu$, but grows most of the time. Moreover, although Gauss law is unambiguous, there is actually not a unique way to write Eq. (\ref{eq1}) in the form Eq. (\ref{eq2}).  However, because the transition to the regime where $\nu \approx 0$ is quite abrupt, there is actually very little freedom in the choice for $\nu$, $v_g$ and $\dRv$ in Eq. (\ref{eq2}), which vindicates the use of that equation and the values we derive for its coefficients. 

The present paper, which is mainly devoted to the derivation of $\I$ and of the envelope equation Eq. (\ref{eq2}), is organized as follows. For pedagogical reasons, we will first present in Section \ref{exp} the derivation of $\I$ in case of a purely time growing wave, and will explain how $\nu$, $v_g$ and $\dRv$ can be deduced from $\I$.  In Section \ref{time}, we will explain how these results can be generalized to a wave whose amplitude either grows or decays in time. Section \ref{space} addresses the issue of a time and space varying wave amplitude, and shows comparisons between our theoretical predictions and results inferred from one dimensional (1-D) simulations of SRS. In this Section will also be discussed how (3-D) effects may affect the range of the validity of the linear regime in terms on the EPW amplitude. In Section \ref{omega} we briefly recall results from Ref.{ }Ê\cite{benisti08} on the nonlinear frequency shift of a driven plasma wave, from which the dephasing $\delta \varphi$ stems and, in Section \ref{srs}, we show one example of the application of our theory to stimulated Raman scattering. Section \ref{conclusion} concludes and summarizes this paper. 

\section{Envelope equation and nonlinear Landau damping rate for a time growing driven plasma wave }
\label{exp}

\begin{figure}[!h]
\centerline{\includegraphics[width=18cm]{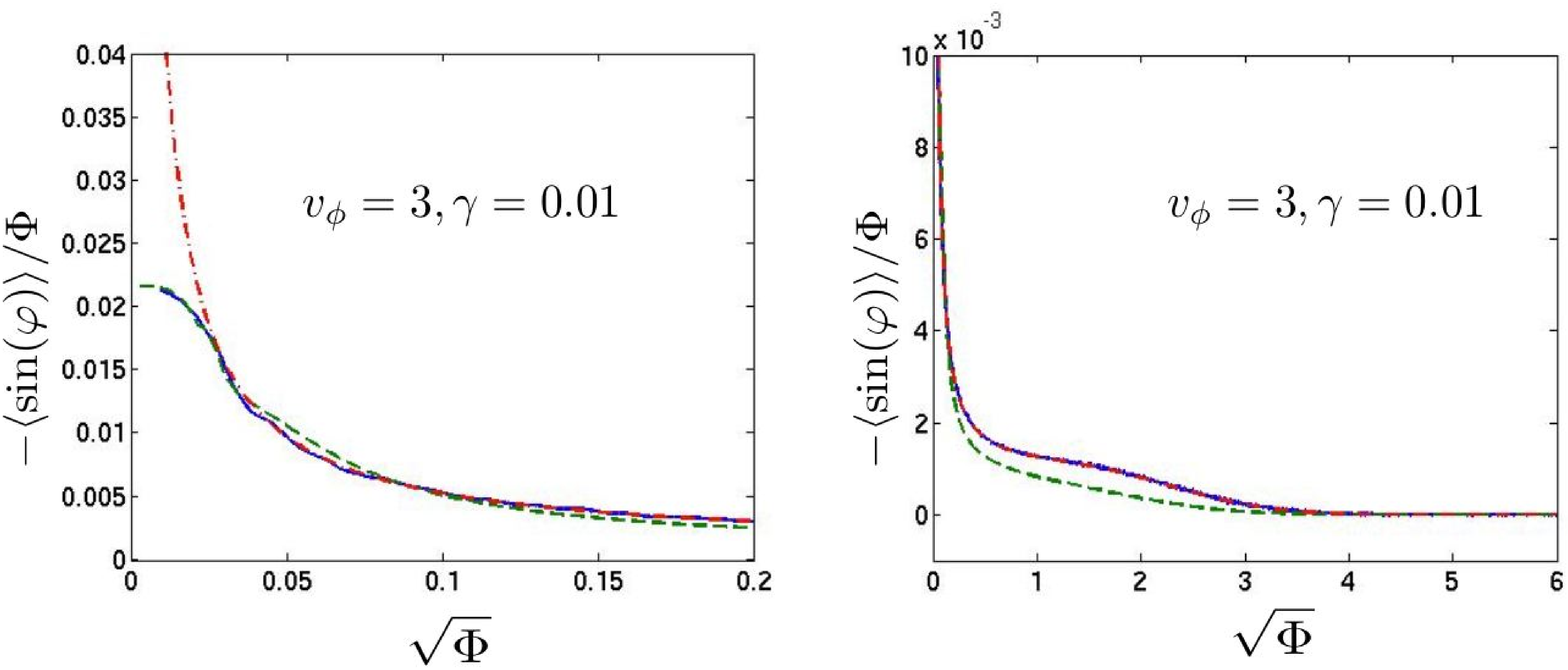}}
\caption{\label{f3} $-\sina/\Phi$ as a function of $\sqrt{\Phi}$ calculated numerically (blue solid line), pertubatively (green dashed line), and adiabatically (red dashed-dotted line)when the normalized wave phase velocity is $v_\phi=3$ and the normalized growth rate is $\gamma=0.01$.}
\end{figure}

\begin{figure}[!h]
\centerline{\includegraphics[width=12cm]{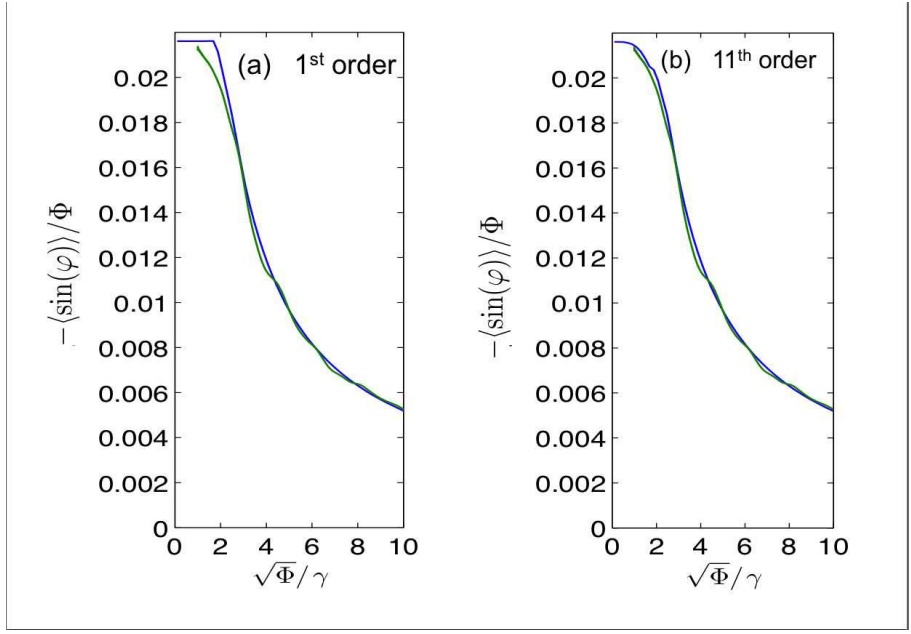}}
\caption{\label{f4} $-\sina/\Phi$ as a function of $\sqrt{\Phi}/\gamma$ when the normalized wave phase velocity is $v_\phi=3$ and the normalized growth rate is $\gamma=0.01$, as calculated numerically (green solid line) and theoretically using Eq. (\ref{eq190}) (blue solid line) with, panel (a), $\text{Im}(\chi_{\text{per}})$ calculated using a $1^{\text{st}}$ order perturbation analysis, panel (b), $\text{Im}(\chi_{\text{per}})$ derived from an $11^{\text{th}}$ order perturbation theory.}
\end{figure}

\begin{figure}[!h]
\centerline{\includegraphics[width=12cm]{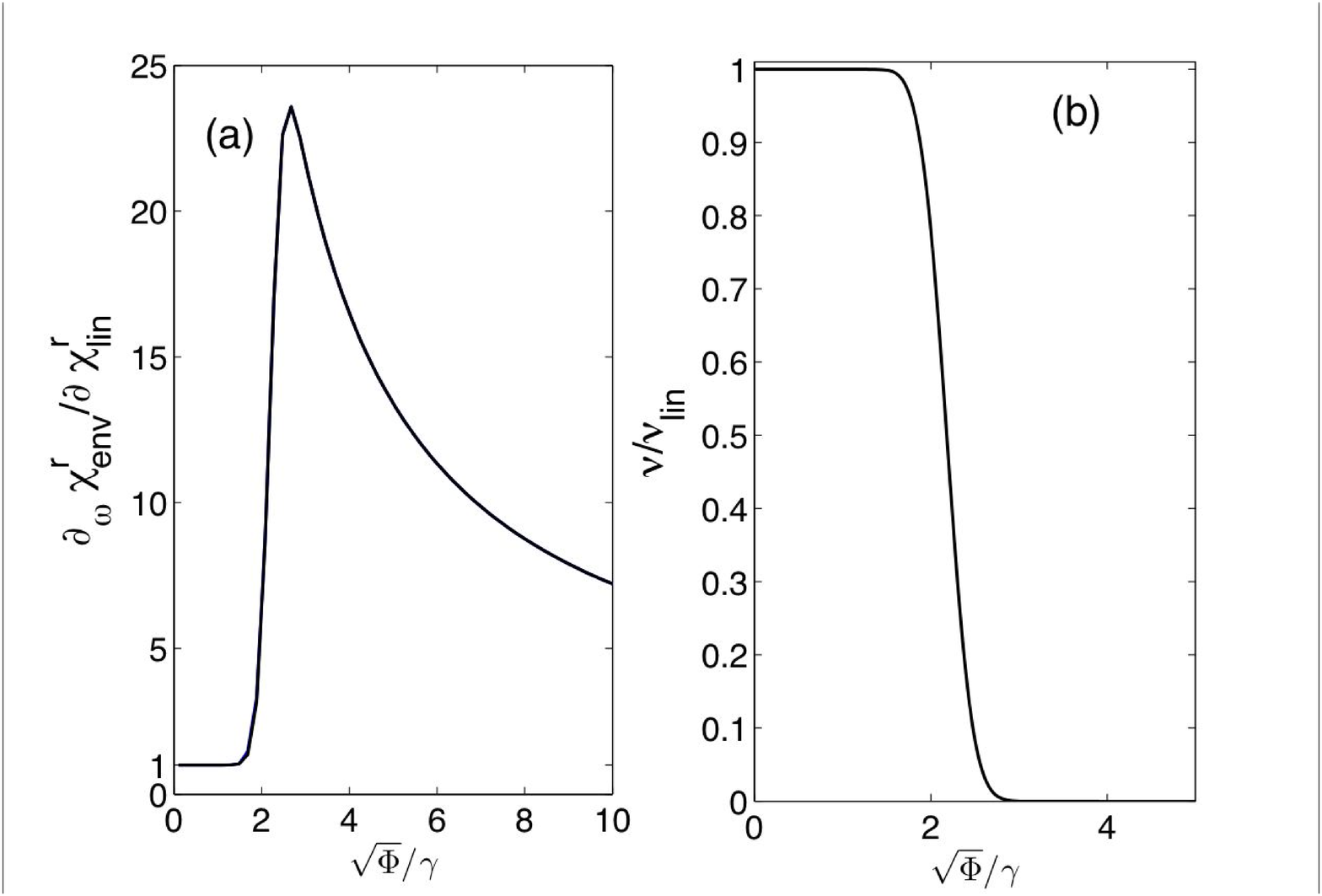}}
\caption{\label{f5} Panel (a), variations of $\dRv$ and, panel (b), variations of $\nu$, normalized to their linear values, as a function of $\sqrt{\Phi}/\gamma$, calculated when $v_\phi=3$ and $\gamma=0.01$ by using a first order perturbative analysis to derive $\text{Im}(\chi_{\text{per}})$.}
\end{figure}

In this Section, we derive the envelope equation for an EPW whose amplitude only depends on time, and grows with time. This will allow us to introduce in a simple way most of the concepts useful in the general situation of a time and space dependence of the wave amplitude. Most of this Section is devoted to the derivation of $\I$, performed by using two very different methods yielding values of $\I$ which do match over a finite range of wave amplitudes. For small amplitudes, we use a perturbative analysis which provides an expression for $\I$ that clearly shows how $\nu$ decreases as more and more electrons are getting trapped in the wave trough. Then, when $\nu \approx 0$, one can approximate $\I$ by, $\I=\Gamma_p \dRv$, where $\Gamma_p$ is the wave growth rate, $\Gamma_p \equiv E_p^{-1} dE_p/dt$, and $\dRv$ is calculated by making use of the adiabatic approximation. As is illustrated in Fig. \ref{f3}, the ``adiabatic'' and perturbative estimates of $\I$ assume very close values over a finite range of wave amplitudes, which allows us to derive an expression for $\I$ valid whatever the wave amplitude by ``connecting'' the two previous estimates, as shown in Fig. \ref{f4}. This connecting is made through a Heaviside like function, leading to abrupt changes in the coefficients of the envelope equation, Eq. (\ref{eq2}). In particular, $\nu$ is found to assume nearly constant values before abruptly dropping to 0, and this drop is concomitant with a sudden rise in $\dRv$ (see Fig. \ref{f5}). Indeed, as will be shown here, that part of $\I$ which, in the linear regime, provides $\nu$, renormalizes $\dRv$ when $\nu \approx 0$.  

\begin{figure}[!h]
\centerline{\includegraphics[width=10cm]{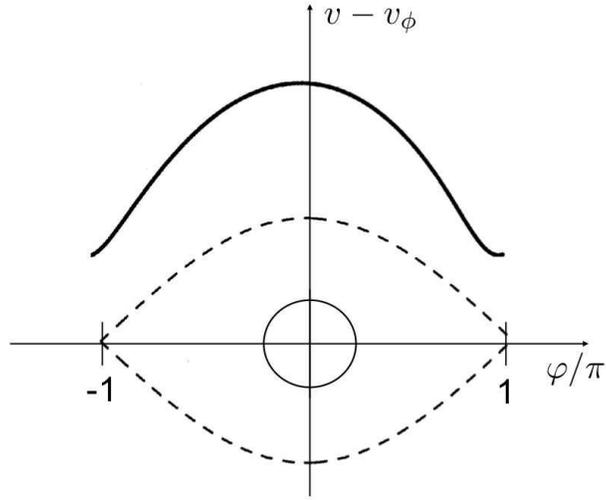}}
\caption{\label{f2} Orbits of electrons acted upon by the longitudinal field $2E_0 \sin(\varphi)$, whose amplitude slowly varies with time. The dashed curve is the virtual separatrix.}
\end{figure}

Let us now enter the details of the theory. We consider here a driven plasma wave, meaning that the total longitudinal field (along the direction of the wave propagation) is the sum of the  EPW field, which is a genuine electrostatic field induced by charge separation, and of the driving field (the so-called ponderomotive field in case of laser drive). We assume that both the electrostatic, $E_{el}(x,t)$, and the driving, $E_{drive}(x,t)$, fields can be expressed in terms of a slowly varying envelope and an eikonal, that is,
\begin{eqnarray}
\label{eq3}
E_{el}(x,t) &\equiv & E_p(t) \sin[\varphi_p(x,t)], \\
\label{eq4}
E_{drive}(x,t)&  \equiv & E_d (t)\cos[\varphi_p(x,t)+\delta \varphi(x,t)],
\end{eqnarray}
with $\vert E_{p,d}^{-1}\partial_x E_{p,d} \vert  \ll \vert k \vert$, $k \equiv \partial_x \varphi_p$, $\vert E_{p,d}^{-1}\partial_t E_{p,d} \vert \ll \vert \omega \vert$, $\omega \equiv -\partial_t \varphi_p$, and $\vert \delta \varphi \vert \ll \vert \varphi_p \vert$. Then, the total longitudinal electric field, including the plasma wave and the drive, also writes in terms of a slowly varying envelope and an eikonal,
\begin{equation}
\label{eq5}
E_{el}+E_{drive}Ê\equiv E_0(t) e^{i\varphi(x,t)}+c.c.,
\end{equation}
where $E_0$ and $\varphi$ are given in terms of $E_p$, $E_d$, $\varphi_p$ and $\delta \varphi$ in Ref. \cite{benisti07}. This total field induces a charge density which may therefore be written as,
\begin{equation}
\label{eq5b}
\rho (x,t) \equiv \rho_0(t) e^{i\varphi}+c.c.
\end{equation}
Throughout this paper we assume immobile ions, and define the electron susceptibility as,
\begin{equation}
\label{eq6}
\chi \equiv \frac{i \rho_0}{\varepsilon_0 k E_0}.
\end{equation}
When the plasma wave is not driven and $E_0$ is an electrostatic field, then Gauss law straightforwardly yields the usual dispersion relation $1+\chi=0$. In the general case, we use the \textit{total} field amplitude $E_0$ in the definition of $\chi$ so that the expression of the electron susceptibility would be the same, in terms of the field amplitude and of the unperturbed distribution function, whether the wave is driven or not. In particular, it is easy to show that, in the linear limit,  $\chi$ is nothing but the usual linear electron susceptibility, as derived in Ref. \cite{fried}. Plugging Eq. (\ref{eq6}) into Gauss law one easily finds,  
\begin{equation}
\label{eq7}
\I E_p=E_d \cos(\delta \varphi),
\end{equation}
provided that $\R \approx -1$ and $\vert \I \vert \ll 1$ (see Ref. \cite{ benisti07} { }Êfor details). In order derive $\I$ and cast Eq. (\ref{eq7}) in the form of the envelope equation, 
\begin{equation}
\label{eq2b}
\partial_t E_p+\nu E_p=E_d \cos (\delta \varphi)/\dRv,
\end{equation}
we now need to express $\chi$ in terms of the electron distribution function. From Eq. (\ref{eq6}), it is clear that $\rho_0$ is nothing but a Fourier component of $\rho$ so that, 
\begin{eqnarray}
\nonumber
\rho_0&=&(2\pi)^{-1}\int_{-\pi}^{\pi}\rho e^{-i\varphi} d\varphi \\
\nonumber
&&= \frac{-ne}{2\pi} \int_{-\pi}^{\pi}\int_{-\infty}^{+\infty}f(\varphi,v,t) e^{-i\varphi} dv d\varphi  \\
\label{eq8}
&& \equiv -ne \langle e^{-i\varphi} \rangle
\end{eqnarray}
where $n$ is electron density, $f$ is the electron distribution function normalized to unity, and $\langle . \rangle$ stands for a local, in space, statistical averaging. For the sake definiteness, and without loss of generality, we henceforth assume that $E_0$ is a pure imaginary number, so that $\I$ is proportional to $\langle \sin(\varphi) \rangle$. As a first step to calculate $\langle \sin(\varphi) \rangle$, we need to evaluate which electrons significantly contribute to it. This is done by investigating the electrons orbits in phase space, schematically displayed in Fig. \ref{f2}. If $E_0$ were a constant, these orbits would be exactly symmetric with respect to the velocity axis, and $\sina$ would be 0. Since $E_0$ slowly varies with time, the electrons orbits are slightly non symmetric, all the more as the growth rate of the total field, $\Gamma \equiv E_0^{-1} dE_0/dt$, is small compared to the time it takes for $\varphi$, or the polar angle in phase space, to change by $2\pi$. This time, henceforth termed the pseudo period of the orbit, is very close to $2\pi/\omega_B$ for a trapped orbit far enough from the virtual separatrix (which is defined by freezing the wave amplitude). Hence, as shown in Fig. \ref{f2}, when $\omega_B \gg \Gamma$ deeply trapped orbits are nearly symmetric with respect to the $v$-axis, and electrons on such orbits contribute very little to $\sina$, and therefore to $\I$. This lets us derive a specific criterion as regards the electrons which need to be accounted for when calculating $\I$. Since $E_0$ varies slowly with time, we use the adiabatic approximation to find out which electrons are trapped in the wave trough. In terms of the dimensionless wave amplitude  $\Phi \equiv eE_0/kT_e$, where $T_e$ is the electron temperature, and of the electron initial velocity, $v_0$, and wave phase velocity, $v_\phi$, both normalized to the thermal velocity, $v_{th} \equiv \sqrt{T_e/m}$, the condition for trapping derived from the adiabatic approximation is, $\vert v_0-v_\phi \vert <4\sqrt{\Phi}/\pi$ (see Ref. \cite{benisti07}). Then, an electron orbit will be considered as ``deeply trapped'' if $\vert v_0-v_\phi \vert <4\sqrt{\Phi}/\pi(1-\delta V)$, with $\delta V$ large enough for the electron orbit to be nearly symmetric. Since the symmetry of a trapped orbit is governed by $\Gamma/\omega_B$, we choose $\delta V$ proportional to $\gamma/\sqrt{\Phi} \equiv \Gamma/\omega_B$ where, in order to stick to dimensionless variables, we have defined $\gamma=\Gamma/kv_{th}$. Therefore, we will henceforth assume that an orbit is deeply trapped, and that the electrons lying on it contribute very little to $\I$, if $\vert v_0-v_\phi \vert < \Vl$, with  $\Vl \equiv \max\left\{0,(4\sqrt{\Phi}/\pi)\left[1-3\gamma/2\sqrt{\Phi} \right] \right\}$, where the value 3/2 has been found numerically (see Ref. \cite{benisti07}). 
\begin{figure}[!h]
\centerline{\includegraphics[width=16cm]{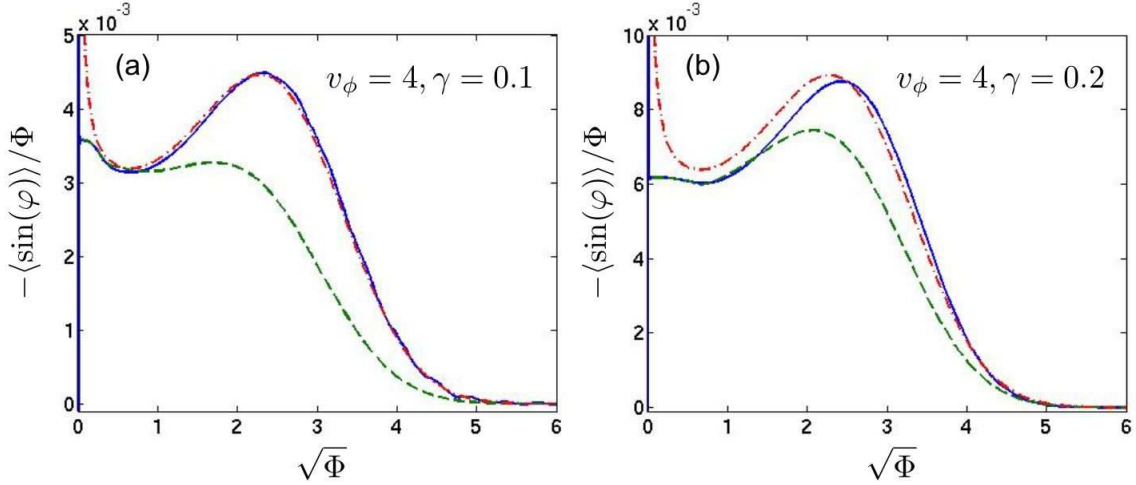}}
\caption{\label{f3b} $-\sina/\Phi$ as a function of $\sqrt{\Phi}$ calculated numerically (blue solid line), pertubatively (green dashed line), and adiabatically (red dashed-dotted line), when the normalized wave phase velocity is $v_\phi=4$ and, panel (a), when the normalized growth rate is $\gamma=0.1$, panel (b), when the normalized growth rate is $\gamma=0.2$.}
\end{figure}
It is noteworthy that, for a slowly growing wave, $\gamma/\sqrt{\Phi}\equiv\Gamma/\omega_B \approx 2/\int \omega_B dt$, so that $\Vl>0$ when $\int \omega_B dt \agt \pi$, that is after the first  trapped electrons have completed about one half of their pseudo periodic orbit and the phase mixing process, introduced by O'Neil in Ref. \cite{oneil}Ê\ to explain the nonlinear decrease of $\nu$, has started to be effective. 

Let us now explain how we actually calculate $\I$ from the matching of two different estimates. For small wave amplitudes (and more precisely when $\sqrt{\Phi} \ll \gamma$), we use a perturbative analysis to derive $\I$, while when $\sqrt{\Phi} \gg \gamma$ (or more precisely $ \Vl\gg \gamma$) we will show that $\I$ is nearly proportional to $\gamma$ and can be very accurately estimated by using the adiabatic approximation. Let us start with the perturbative estimate of $\I$. There are several reasons to believe that a perturbative analysis will be useful in deriving $\I$. First, it has been proven in Ref. \cite{villani} that for small enough wave amplitudes linear theory, which stems from a first order perturbative analysis of the electron motion, is valid. Second, as shown before, one may neglect the contribution of the deeply trapped electrons,  whose motions cannot be treated perturbatively, to estimate $\I$. Mathematically, this amounts to bounding from below the small denominators in the perturbative expression of $\I$.  Actually, although rigourous estimates remain to be done, it appears from the results of Ref. \cite{benisti07} that the ``small parameter'' of the perturbative expansion of $\I$ varies from $\sqrt{\Phi}/\gamma$  when $\Vl \ll \gamma$, to $\sqrt{\Phi}/\Vl$ when $\Vl \gg \gamma$, and hence remains bounded. However, a perturbative estimate of $\I$ eventually ceases to be accurate as the wave grows.  Physically, this may be understood by the fact that, as $\sqrt{\Phi}/\gamma$ increases, the electrons have to lie on orbits closer to the separatrix to significantly contribute to $\I$, and the motion close to the separatrix is known to be non perturbative. Note, again, that $\sqrt{\Phi}/\gamma \approx \int \omega_B dt /2$, so that $\sqrt{\Phi}/\gamma$ be large corresponds to the usual criterion for a highly nonlinear, and hence non perturbative, electron response. 

Let us now detail the perturbative expression of $\I$, which sheds a lot of light on the nonlinear decrease of $\nu$, and on how $\I$ may be estimated when $\sqrt{\Phi}/\gamma$ is large. Indeed, at first order in the perturbation analysis, and at 0-order in the time variations of $\gamma$ and $v_\phi$, one finds the well known result (see Ref. \cite{benisti07} for details),
\begin{equation}
\label{o1}
\I=\frac{-2}{(\kld)^2}\int_{\vert v \vert>\Vl} \frac{\gamma v}{(\gamma^2+v^2)^2} f_0(v+v_\phi)dv,
\end{equation}
where $\lambda_D\equiv v_{th}/\wpe$ is the Debye length and $f_0$ is the normalized electron distribution function in the limit of a vanishing field amplitude. Now, Eq. (\ref{eq7}) derived from Gauss law is the envelope equation Eq. (\ref{eq2b}) only if $\I$ may be written as, $\I \approx \delta I_1+\Gamma_p \delta I_2$, where $\Gamma_p \equiv E_p^{-1} dE_p/dt$, and where $\delta I_1$ and $\delta I_2$ only depend on the wave amplitude and not on its time variations (at least over finite ranges of amplitudes). As shown in Ref. \cite{benisti08}, $E_p \gg E_d$, so that $E_p\approx E_0$, except if $\nu$ is so large that the term $\nu E_p$ dominates in the left hand side of Eq. (\ref{eq2b}) and $E_p \approx E_d/\nu$. Hence, $\Gamma_p \approx \Gamma \equiv E_0^{-1} dE_0/dt$. We therefore only need to write $\I$ as $\I \approx \delta I_1+\Gamma \delta I_2$, which clearly requires to isolate the divergence of the integrand in Eq. (\ref{o1}) when $\Vl=0$ and $\gamma \rightarrow 0$. We do this by using the following decomposition, $\I=I_1+I_2$, with,
\begin{eqnarray}
\nonumber
I_1 &\equiv& \frac{-2f'_0(v_\phi)}{(k\lambda_D)^2} \int_{\vert v \vert>\Vl} \frac{\gamma v^2}{(\gamma^2+v^2)^2}dv \\
\label{I1}
&=&-\frac{f'_0(v_\phi)}{(\kld)^2}\left[\pi-2\tan^{-1}\left(\frac{\Vl}{\gamma}\right) +\frac{2\gamma \Vl}{\gamma^2+\Vl^2}\right], \\
\label{I2}
I_2&\equiv&\frac{-2\gamma}{(k\lambda_D)^2}  \int_{\vert v \vert>\Vl} \frac{v}{(\gamma^2+v^2)^2} [f_0(v+v_\phi)-vf'_0(v_\phi)]dv.
\end{eqnarray}
Since $\gamma \ll 1$, one may approximate $I_2$ by replacing $(\gamma^2+v^2)$ by $v^2$ to find,
\begin{eqnarray} 
\label{eq18}
I_2& \approx &\frac{-2}{(\kld)^2} \gamma  \int_{\vert v \vert > \Vl} \frac{f_0(v+v_\phi)-vf'_0(v_\phi)}{v^3}dv \\
\label{eqx}
& \equiv &  \Gamma (\partial \chi_1^r/\partial \omega),
\end{eqnarray}
where the integral in Eq. (\ref{eq18}) has to be taken in the sense of Cauchy's principal part when $\Vl=0$.

When $\Vl=0$, $\chi_1^r$ is just the adiabatic approximation of the linear value of $\text{Re}(\chi)$ (and its value does not change much provided that $\Vl \alt 1$), while $I_1=-\pi (\kld)^{-2}f'_0(v_\phi)$. Hence, $\I$ is in the desired form, $\I=-\pi (\kld)^{-2}f'_0(v_\phi) +\Gamma \partial_\omega \chi_1^r$, so that Eq. (\ref{eq7}) may indeed be written as the envelope equation, Eq. (\ref{eq2b}), with $\dRv= \partial_\omega \chi_1^r$, and $\nu=\nu_L$, the Landau damping rate in the limit $\nu_L \ll \wpe$. 

When $\Vl \gg \gamma$, $I_1$ is nearly proportional to $\gamma$ and therefore so is $\I$, which is actually obvious from Eq. (\ref{o1}). Then, Eq. (\ref{eq7}) straightforwardly writes as Eq. (\ref{eq2b}) with $\nu\approx 0$; Landau damping has become negligible in the time evolution of the driven plasma wave.  Physically, the decrease of $\nu$ towards 0 is due to the trapping of the nearly resonant electrons, which no longer contribute to $\nu$ while oscillating in the wave trough, just  like in the situation considered by O'Neil. 

Replacing $(\gamma^2+v^2)$ by $v^2$ in Eq. (\ref{o1}), which is valid when $\Vl \gg \gamma$, one actually finds $\I = \Gamma (\partial_\omega \Rf)^1$, where $(\Rf)^1$ is the first order estimate of some effective real susceptibility, calculated adiabatically and by removing the contribution of the deeply trapped electrons.  There is however no need to resort to perturbation analysis to calculate $\Rf$ since this can be done by directly using the adiabatic approximation, as shown in Ref. \cite{benisti07}. Then, from the previous discussion, we expect that when $\sqrt{\Phi} \gg \gamma$, $\I \approx \Gamma \partial_\omega \Rf$, which provides a non perturbative estimate of $\I$, which will henceforth be termed the ``adiabatic estimate'' of $\I$ [although this is not a proper terminology since a direct adiabatic calculation of $\I$ would just yield $\I=0$]. It is noteworthy that the $I_1$ term Eq. (\ref {I1}) which, in the linear limit provides $\nu$, fully contributes to $\partial_\omega \Rf$ when $\nu \approx 0$.  In the strong damping limit, when $\nu_L \gg \Gamma$, $\dRv$ may then increase by more than one order of magnitude, as illustrated in Fig. \ref{f5}.

Let us now compare the perturbative and adiabatic estimates of $\I$ to those derived from test particles simulations. Numerically, we calculate the dynamics of electrons acted upon by an exponentially growing wave and estimate $\sina=\sum_{i=1}^N w_i \sin(\varphi_i)$, where the sum runs over all the electrons used in the simulation, and $w_i \equiv f_0(v_{0i})$, where $v_{0i}$ is the initial velocity of the i$^{\text{th}}$ electron and $f_0$ is the normalized unperturbed distribution function. In our simulations, we chose $f_0(v)=(2\pi)^{-1/2}\exp(-v^2/2)$. Whatever the wave phase velocity and for small enough growth rates, we always found that the high ($11^{\text{th}}$) order perturbative estimate of $\I$ was valid at least up to $\sqrt{\Phi}/\gamma \approx 10$, while the adiabatic estimate was correct whenever $\sqrt{\Phi}/\gamma \agt 3$ (see Fig. \ref{f3}). Such comparisons moreover allowed us to conclude that an adiabatic estimate of $\I$ was only accurate if $\gamma \alt 0.1$, as illustrated in  Fig. \ref{f3b}. 

Using the perturbative, $\text{Im}(\chi_{\text{per}})$, and adiabatic estimates of $\I$ within their respective ranges of validity, which do overlap, we obtain the following expression for $\I$, valid whatever the wave amplitude,
\begin{equation}
\label{eq190}
\I =  \text{Im}(\chi_{\text{per}})\left[1- Y\left(\sqrt{\Phi}/\gamma \right)\right]+\Gamma \partial_\omega\Rf Y\left(\sqrt{\Phi}/\gamma \right),
\end{equation}
where $Y$ is a function rising from 0 to 1 as $\sqrt{\Phi}/\gamma$ increases. Since, as shown in Fig. \ref{f3}, the convergence of $\Gamma \partial_\omega\Rf$ towards $\I$ is quite sharp, $Y$ should rise very quickly from 0 to 1 as $\sqrt{\Phi}/\gamma$ increases from a little less than 3 to a little more than 3. This is the case if we choose $Y(x)=\tanh^5[(e^{x/3}-1)^3]$. Fig. \ref{f4} shows comparisons between theoretical values of $-\sina/\Phi$ derived from Eq. (\ref{eq190}), and numerical ones provided by test particles simulations. From this Figure, it is clear that  using a high (11$^{\text{th}}$) order perturbative expression for $\text{Im}(\chi_{\text{per}})$ yields very accurate values for $-\sina/\Phi$, and hence for $\I$, while calculating $\text{Im}(\chi_{\text{per}})$ at first order already yields very good results, with much more simple formulas! Therefore, for practical purposes such as the numerical study of SRS, we restrict to first order expressions.  Then, from Eq. (\ref{eq190}) and the expression found previously for $\text{Im}(\chi_{\text{per}})$, we conclude that Gauss equation, Eq. (\ref{eq7}), is the envelope equation, Eq. (\ref{eq2b}), with,
\begin{eqnarray}
\label{eq20}
\chi^r_\text{env}&=&(1-Y)\times \chi_1^r+Y\times \chi^r_{\text{eff}}, \\
\label{eq21}
\nu&=&Y \times I_1/\dRv \approx Y \times I_1/\partial_\omega \chi_1^r,
\end{eqnarray}
where $I_1$ and $\chi_1^r$ are defined by Eqs. (\ref{I1}) and (\ref{eqx}). If we were to replace $\gamma$ by $(k v_{th}ÊE_p)^{-1} dE_p/dt$ in the expression (\ref{I1}) for $I_1$, we would find that $\nu$ actually is much more complicated an operator than a plain damping rate. However, as shown in Fig. \ref{f5}, provided that $\gamma$ remains nearly constant, $\nu$ assumes nearly constant values before abruptly dropping to 0. $\nu$ may then indeed be viewed as a damping rate, both physically and when numerically solving the envelope equation, Eq. (\ref{eq2b}). We therefore successfully defined an effective nonlinear damping rate, $\nu$, yielding the time evolution of the driven plasma wave, which was our prime goal. We term $\nu$ the ``nonlinear Landau damping rate'' of the driven plasma wave because it physically stems from the electron acceleration by the EPW, which is the very mechanism giving rise to the Landau damping of a freely propagating wave. Then, as expected, the linear value of $\nu$ is nothing but the Landau damping rate. Note that we relate $\nu$ to the growth of the driven plasma wave and not to any other quantity, such as the energy gain by the electrons from the wave. As shown in Fig. \ref{f5}, the drop in $\nu$ is concomitant with a rapid growth of $\dRv$, since the term in $\I$ which gives rise to $\nu$ in the linear regime fully contributes to $\dRv$ when $\nu \approx 0$, so that $\I$, and the efficiency of the driving of the EPW, varies smoothly. 

\section{Generalization to an arbitrary time dependence of the wave amplitude}
\label{time}
In this Section, we generalize the results derived previously to a plasma wave whose amplitude may vary arbitrarily in time, provided that the growth rate, $\Gamma \equiv E_0^{-1}dE_0/dt$, is still such that $\vert \Gamma \vert \ll \wpe$. We shall moreover show that the formula (\ref{eq21}) for $\nu$, with $I_1$ given by Eq. (\ref{I1}), is still useful provided that $\gamma$ is defined properly \textit{i.e.},  by Eq. (\ref{eq111}).

\begin{figure}[!h]
\centerline{\includegraphics[width=14cm]{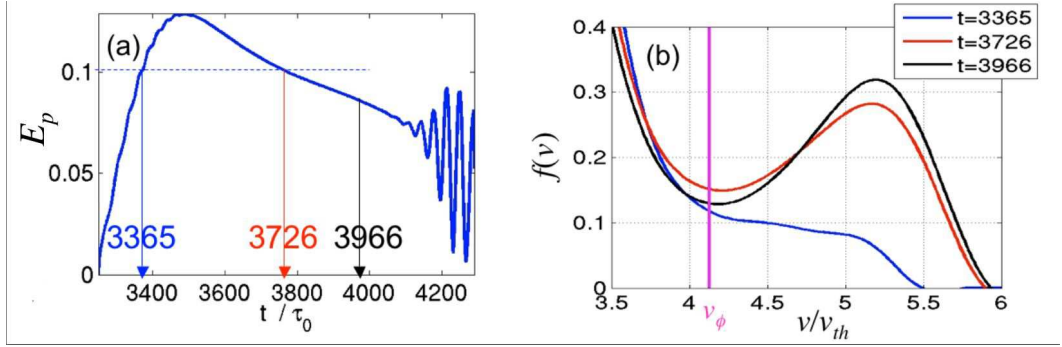}}
\caption{\label{f6} Results from Valsov simulations of stimulated Raman scattering showing, panel (a), the plasma wave amplitude (in its reference frame and in arbitrary units) as a function of time (normalized to the laser period), panel (b), the space averaged electron distribution function at the three different times indicated by the arrows in panel (a). Note that, as the EPW amplitude decreases, the space averaged distribution function becomes more symmetric with respect to $v_\phi$, and that it is not only a function of the EPW amplitude. }
\end{figure}

We start by estimating $\langle e^{-i\varphi}Ê\rangle$ through the means of a first order perturbation analysis, which proved in the preceding Section to be an important step in the derivation of $\I$. By using the Hamiltonian perturbation analysis detailed in Appendix \ref{A}, one finds, at first order, $\varphi(\tau)=\varphi_0+(v_0-v_\phi)\tau+\delta \varphi$, where $\tau=k\lambda_D \omega_{pe}t$, velocities are still normalized to the thermal one, and,
\begin{equation}
\delta \varphi = -i e^{i(\varphi_0+w\tau)}\frac{\partial}{\partial w}\int_0^\tau \Phi(u)e^{iw(u-\tau)}du+c.c.
\end{equation}
where we have denoted $w\equiv v_0-v_\phi$. Then, 
\begin{eqnarray}
\nonumber
\langle e^{-i\varphi} \rangle &\approx& \langle-i \delta \varphi e^{-i(\varphi_0+w\tau)} \rangle\\
\label{eq101}
&=&-\int_{\vert w \vert >\Vl} f_0(w+v_\phi)\frac{\partial}{\partial w} \int_0^\tau \Phi(u) e^{iw(u-\tau)}du dw,
\end{eqnarray}
where $\Vl$ is a straightforward generalization, using the phase mixing argument, of the value found in the previous Section \textit{i.e.}, $\Vl = 4\sqrt{\Phi}/\pi\left[1-3/\int_0^t\omega_B(u)du\right]$, and where $f_0$ is the electron distribution function in the limit $\Phi \rightarrow 0$. If $\Phi$ has kept on increasing with time, $f_0$ is nothing but the unperturbed distribution function. If $\Phi$ has reached a large enough value to induce nonlinear electron motion before decreasing to nearly 0, a perturbative analysis of the electron motion from $t=0$ is no longer valid. However, one may calculate the electron motion perturbatively from $t=+\infty$ by making use of the time-reversal invariance of the dynamics. Then, $f_0$ is the distribution function in the limit $t \rightarrow +\infty$ which, as shown in Ref. \cite{benisti07}, and as illustrated in Fig. \ref{f6}, results from symmetric detrapping. In the interval $\vert v-v_\phi \vert >\max(\Vl)$, $f_0(v,t=+\infty)$ assumes the same values as the initial, unperturbed, distribution function, while in the interval $\vert v-v_\phi \vert \leq \max(\Vl)$, $f_0(v,t=+\infty)$ is nearly symmetric with respect to $v_\phi$. Then, electrons whose initial velocity lies within the latest interval contribute very little to $\I$. This means that once deeply trapped, electrons no longer contribute significantly to $\I$, even after being detrapped. Eq. (\ref{eq101}) may therefore be simplified by using for $f_0$ the unperturbed distribution function and by replacing $\Vl$ by $\max(\Vl)$. Such a simplfication will be implicitly used throughout the remainder of this paper.

We now use the same kind of decomposition as in the previous Section to find a suitable expression of $\I$ \textit{i.e.}, we write, $\chi=-i(\kld)^{-2}\langle e^{-i\varphi} \rangle/\Phi \equiv \chi_a+\chi_b$, with,
\begin{eqnarray}
\label{E1}
\chi_a&=&\frac{if'_0(v_\phi)}{(\kld)^2\Phi(\tau)}\int_0^\tau \Phi(u)(u-\tau) \int_{\vert w \vert >\Vl} iw e^{iw(\xi-\tau)}dw du, \\
\nonumber
\chi_b&=&\frac{i}{(\kld)^2\Phi(\tau)}\int_{\vert w \vert >\Vl} [f_0(w+v_\phi)-wf'_0(v_\phi)] \times \\
\label{E2}
&&\left(\frac{\partial}{\partial w} \int_0^\tau \Phi(u)e^{iw(u-\tau)} du \right) dw.
\end{eqnarray}
Provided that $\Phi(\tau) \gg \Phi(0)$, integrating Eq. (\ref{E2}) by parts with respect to time yields, at first order in the time variations of $\Phi$,
\begin{eqnarray} 
\label{eq102}
\text{Im}(\chi_b)& \approx &-2 (\kld)^{-2}\Phi^{-1}\frac{d\Phi}{d\tau}  \int_{\vert w \vert > \Vl} \frac{f_0(w+v_\phi)-wf'_0(v_\phi)}{w^3}dw \\
\label{eq103}
& \equiv &\Gamma (\partial \chi_1^r/\partial \omega)
\end{eqnarray}
where, again, the integral in Eq. (\ref{eq103}) has to be taken in the sense of Cauchy's principal part when $\Vl=0$. Hence, in the limit of a slowly varying wave amplitude, the expression of $\text{Im}(\chi_b)$ is exactly the same as that of the term $I_2$ found in the previous Section, Eq. (\ref{eq18}).

When $\Vl=0$, since $\int_{-\infty}^{+\infty} iw e^{iw(u-\tau)}dw=2\pi \partial_u \delta(u-\tau)$, where $\delta$ is the Dirac distribution, one easily finds $\text{Im}(\chi_a)=-(\kld)^2\pi f'_0(v_\phi)$. Hence when $\Vl=0$, which corresponds to the linear limit,  $\I=\Gamma \partial_\omega \chi_1^r-\pi (\kld)^{-2} f'_0(v_\phi)$, so that Gauss equation (\ref{eq7}) is,
\begin{equation}
\partial_t E_p+\nu_L E_p=E_d \cos(\delta \varphi)/\partial_\omega \chi_1^r,
\end{equation}
where $\nu_L$ is the Landau damping rate, in the limit $\nu_L \ll \wpe$. Since our linear calculation is valid whether the plasma wave is driven, or not, it unambiguously shows Landau damping, without resorting to complex contour deformation. This therefore allows us to conclude that non-Landau damping, as described by Blemont \textit{et al.} in Ref. \cite{belmont}, cannot be obtained by using a drive at the same frequency as the plasma wave to excite it above the noise level, and then let it freely propagate. 

In the nonlinear regime, and when $\Vl^{-1}$ is much smaller than the typical timescale of variation of $\Phi$, $\tau_{\phi}$, calculating the time integral in Eq. (\ref{E1}) by parts yields,
\begin{equation}
\label{eq104}
\text{Im}(\chi_a)=-(\kld)^{-2}\Phi^{-1}f'_0(v_\phi)[4\Vl^{-1} d\Phi/d\tau+O(\Vl^{-3}d^3\Phi/d\tau^3)].
\end{equation}
Hence, when $\Vl \gg \tau_{\phi}^{-1}$, $\Ic$ is nearly proportional to $\Gamma$, and therefore so is $\I$, which implies $\nu \approx 0$. Again, as in the previous Section, we find that the decrease of $\nu$ towards 0 is due to the trapping of the nearly resonant electrons. Moreover, it is easy to show that in the limit $\Vl \gg \gamma$, the $I_1$ term Eq. (\ref{I1}) of the previous Section, is close to $-4(\kld)^{-2}f'_0(v_\phi)\gamma/\Vl$, just as $\text{Im}(\chi_a)$. We therefore conclude that, in the limit of large $\Vl$, when $\nu \approx 0$, the results obtained in the previous Section for a growing wave, are valid whatever the time dependence of the wave amplitude. Hence, when $\Vl \gg \tau_{\phi}^{-1}$ which, for a slowly varying wave is typically the case when $\int_0^t \omega_Bdu \gg 1$, we expect $\I \approx \Gamma \partial_\omega \Rf$, where $\Rf$ is the same as in the preceding Section. Then, generalizing the results of Section \ref{exp}, we propose the following expression for $\I$, 
\begin{equation}
\label{eq105}
\I =  \text{Im}(\chi_{\text{per}})\left[1- Y\left(2\int_0^t \omega_B du \right)\right]+\Gamma \partial_\omega\Rf Y\left(2\int_0^t \omega_B du \right),
\end{equation}
where $Y$ is the same function as for a growing wave, and when $\text{Im}(\chi_{\text{per}})$ is still the perturbative estimate of $\I$ which, at first order, is $\text{Im}(\chi_{\text{per}})=\text{Im}(\chi_a+\chi_b)$ defined by Eqs. (\ref{E1},\ref{E2}). Eq. (\ref{eq105}), when generalized to allow for the space variation of the wave amplitude, yields results in very good agreement with those inferred from Vlasov simulations of SRS, as shown in Fig. \ref{f7}. From the expression (\ref{eq105}) of $\I$, and Gauss law Eq. (\ref{eq7}), we derive the envelope equation (\ref{eq2b}) with,
\begin{eqnarray}
\label{eq106}
\chi^r_\text{env}&=&(1-Y)\times \chi_1^r+Y\times \chi^r_{\text{eff}}, \\
\label{eq107}
\nu&=&Y \times\Ic/\dRv.
\end{eqnarray}
Since, whenever $Y$ is not close to 0,  $\chi^r_\text{env} \approx \chi_1^r$, Eq. (\ref{eq107}) for $\nu$ may be simplified in,
\begin{equation}
\label{eq108}
\nu \approx Y \times\Ic/\partial_\omega \chi_1^r.
\end{equation}

We now try to find a more simple expression for $\Ic$, leading to a practical analytic formula for $\nu$. In the limit of large $\Vl$, we already showed that $\Ic$ was well approximated by Eq. (\ref{I1}) for $I_1$. In the opposite limit when $\Vl \ll \tau_\Phi^{-1}$, as shown in Appendix \ref{B}, we find, $\Ic=-(\kld)^{-2}f'_0(v_\phi)[\pi+\delta \chi_a]$, with,
\begin{equation}
\label{eq109}
\delta \chi_a \approx -\frac{4\Vl^3}{3\Phi(\tau)}\int_0^\tau \int_0^u\int_0^\xi \Phi(\xi')d\xi'd\xi du.
\end{equation}
Similarly, when $\Vl \ll \gamma$, a Taylor expansion of Eq. (\ref{I1}) yields $I_1=-(\kld)^{-2}f'_0(v_\phi)[\pi+\delta I_1]$, with $\delta I_1 \approx -(4/3)(\Vl/\gamma)^3$. Since, for a slowly varying wave, and when $\Phi(\tau) \ggÊ\Phi(0)$, $\delta \chi_a \approx -[4\Vl^3/3\Phi(\tau)] \left( \int_0^\tau \Phi(u) du \right)^3$, we find that Eq. (\ref{I1}) still applies in the general case, and in the limit $\Vl \ll \tau_\Phi^{-1}$, provided that $\gamma$ be replaced by $\Phi(\tau)/\int_0^\tau \Phi(u) du$. Hence, while for an exponential growing wave, for which Eq. (\ref{I1}) is exact, $\gamma \equiv \Phi^{-1}d\Phi/d\tau=\Phi(\tau)/\int_0^\tau \Phi(u) du$, we find that this equation still holds in the general case provided that,  $\gamma= \Phi(\tau)/\int_0^\tau \Phi(u) du$ when $\Vl \ll \tau_\Phi^{-1}$, and $\gamma=\Phi^{-1}d\Phi/d\tau$ when $\Vl \gg \tau_\phi^{-1}$. Therefore, we propose the following approximate expression for $\Ic$,
\begin{eqnarray}
\label{eq110}
\Ic&=&-\frac{f'_0(v_\phi)}{(\kld)^2}\left[\pi-2\tan^{-1}\left(\frac{\Vl}{\gamma}\right) +\frac{2\gamma \Vl}{\gamma^2+\Vl^2}\right], \\
\label{eq111}
\gamma &=&\frac{\Phi(\tau)-\Phi(\tau-\pi/\Vl)}{\int_{\tau-\pi/\Vl}^\tau \Phi(u)du},
\end{eqnarray}
where it is clear that $\gamma$ defined by Eq. (\ref{eq111}) has the required properties, $\gamma \approx \Phi(\tau)/\int_0^\tau \Phi(u) du$ when $\Vl \ll \tau_\Phi^{-1}$, and $\gamma \approx \Phi^{-1}d\Phi/d\tau$ when $\Vl \gg \tau_\phi^{-1}$. Eqs. (\ref{eq110},\ref{eq111}) have been used when comparing our theoretical estimate to numerical ones, and the good agreement between these two estimates, illustrated in Fig. \ref{f7}, shows the relevance of our approximation. Then, Eq. (\ref{eq108}), together with Eqs. (\ref{eq110},\ref{eq111}), provide a practical analytic formula for $\nu$. The accuracy for $\I$, and thus for $\nu$, can even be improved by using, instead of Eq. (\ref{eq110}), a result derived at higher order in the perturbative analysis (see Ref. (\cite{benisti07}). 

\section{Space and time variation of the wave amplitude}
\label{space}
\subsection{One dimensional (1-D) space variation and comparisons with 1-D simulations of Stimulated Raman Scattering}
\label{1D}
\begin{figure}
  \centerline{\includegraphics[width=12cm]{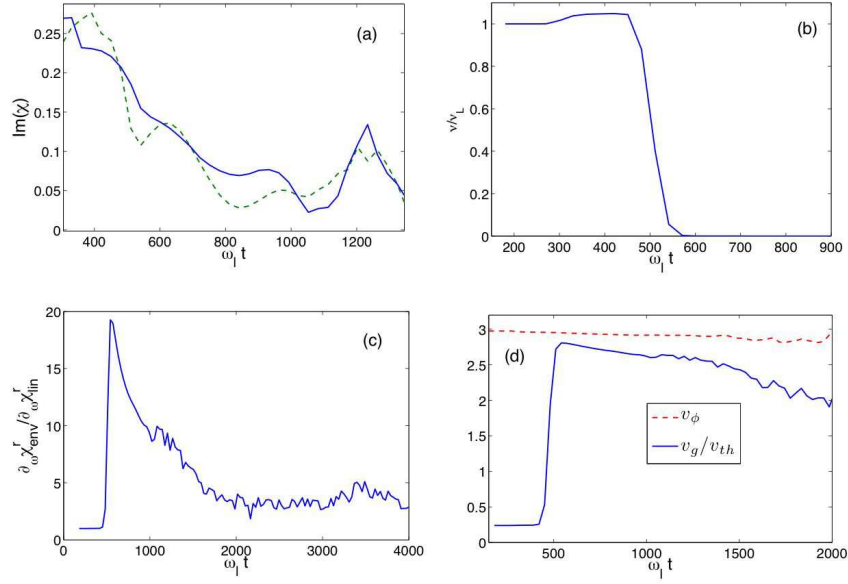}}
   \caption{\label{f1} Panel (a), $\I$ as calculated numerically (blue solid line) and theoretically (green dashed line, panel (b) $\nu$ normalized to the Landau damping rate, panel (c), $\dRv$ normalized to its linear value and, panel (d), the EPW group velocity (blue solid line) and phase velocity (red dashed line) normalized to the thermal one.}
\end{figure}

The results of the previous Section are easily generalized to allow for one dimensional (1-D) space variations of the EPW amplitude.  Indeed, using a Fourier expansion of the charge density, one finds, as shown in Ref. \cite{benisti07}, 
\begin{equation}
\label{eq201}
\I=\text{Im}(\chi_{0D})-\kappa[\partial_k \Rv+\R/k],
\end{equation}
where $\kappa \equiv E_0^{-1}\partial_xE_0 \approx E_p^{-1}Ê\partial_x E_p$, and where $\text{Im}(\chi_{0D})$ is given by Eq. (\ref{eq105}) except that all quantities must now be evaluated in the wave frame. More precisely, $\int_0^t \omega_B du$ in Eq. (\ref{eq105}) or in the definition of $\Vl$ now is,  $\int_0^t \omega_B[x-\int_u^t v_\phi(t')dt',u]du$, and the value for $\gamma$ to be used in Eq. (\ref{eq110}) is,
\begin{equation}
\label{eq202}
\gamma(x,\tau)=\frac{\Phi(x,\tau)-\Phi\left[x-\int_{\tau-\pi/\Vl}^\tau v_\phi(u)du,\tau-\pi/\VlÊ\right]}{\int_{\tau-\pi/\Vl}^\tau \Phi \left[x-\int_u^t v_\phi(t')dt',u\right]du}.
\end{equation}
Plugging Eq. (\ref{eq201}) into Gauss equation (\ref{eq1}), we find the following envelope equation,
\begin{equation}
\label{eq203}
\partial_t E_p+v_g \partial_x E_p+\nu E_p=E_d \cos (\delta \varphi)/\dRv,
\end{equation}
where, provided that $[1+\R] \approx 0$, $v_g = -\partial_k\Rv/\dRv=\omega/k-2/[k\dRv]$. It is noteworthy that, since in the nonlinear regime $\Rv \neq \R$, $v_g \neq d\omega/dk$. Actually, since $\dRv$ may reach values much larger than in the linear limit, the nonlinear value of $v_g$ may get quite close to the EPW phase velocity, as  shown in Fig. \ref{f1} (d). Moreover, $d\omega/dk$ may actually change sign from positive to negative, at rather small wave amplitudes (if $\kld$ is large enough), which would entail a shock in the plasma wave profile if $v_g$ were indeed $d\omega/dk$, while such a shock is not observed in kinetic simulations of Stimulated Raman Scattering. This is an indirect evidence that $v_g \neq d\omega/dk$.

We now compare our theoretical calculations of $\I$ against direct 1-D Vlasov simulations of SRS using the Eulerian code ELVIS \cite{strozzi}. In our numerical simulations, which are detailed in Refs. \cite{strozzi, benisti08}, the EPW results from the interaction of a pump laser, entering from vacuum on the left
\begin{figure}
  \centerline{\includegraphics[width=10cm]{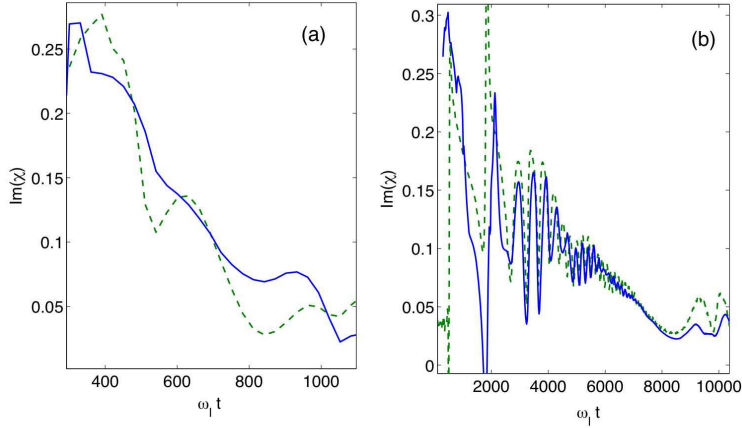}}
   \caption{\label{f7} Time variations of $\I$ as calculated theoretically (green dashed line), and as calculated numerically (blue solid line) without, panel (a), or with, panel (b), using a Lorentzian factor in the $vÊ\times B$ term of Vlasov equation. }
\end{figure}
($x=0$), and a small-amplitude counterpropagating ``seed'' light wave injected
on the right. Using a Hilbert transform of the fields, one can numerically calculate the ratio $[E_d \cos(\delta \varphi)+k^{-1}\partial_xE_p]/E_p$, which from Eq. (\ref{eq1}) yields a first, numerical estimate, of $\I$. From Vlasov simulations one can also extract the values of all the quantities, such as $\int \omega_B dt$ and $\gamma$, which enter our theoretical formula for $\I$. Using these values we calculate a second, theoretical estimate, for $\I$. Both these estimates are compared in Fig. \ref{f7}, plotting $\I$ as a function of $\omega_l t$, where $\omega_l$ is the laser frequency. The simulation results of Fig. \ref{f7} correspond to a plasma with electron temperature, $T_e=5$keV, and electron density $n=0.1 n_c$, where $n_c$ is the critical density. The laser intensity is $I_l=4\times10^{15}$W/cm$^2$ while the seed intensity is $I_s=10^{-5}I_l$ and the seed wavelength is $\lambda_s=0.609\mu m$. The results plotted on Fig. \ref{f7} (a) correspond to a simulation box of length $L=285 \lambda_l$, where $\lambda_l=0.351 \mu m$ is the laser wavelength, and were measured at $x=77 \lambda_l$. In case of Fig. \ref{f7}Ê(b), the length of the simulation box is $L=350 \lambda_l$, while the data were measured at $x=150 \lambda_l$. Moreover, in case of Fig. \ref{f7}Ê(b), the $v\times B$ term in Vlasov equation was artificially multiplied by a Lorentzian factor, so as to mimic laser focusing which would occur in more than one space dimension. As can be seen in Fig. \ref{f7},  there is a very good agreement between the theoretical and numerical values of $\I$, especially as regards the decrease of $\I$ from its linear value in Fig. \ref{f7} (a), while the oscillations in $\I$ due to those of $\gamma$ are very well reproduced in Fig. \ref{f7} (b). 

The time variations of all the terms in the envelope equation (\ref{eq203}) are plotted in Fig. \ref{f1} for the same conditions as in Fig. \ref{f7} (a). Fig. \ref{f1}Ê(b) clearly shows that $\nu$ remains nearly constant before abruptly dropping to 0, and that this is concomitant with a sudden rise in $\dRv$, as for a purely time growing wave. This is very different from the oscillating result found by O'Neil because, in this paper, we consider slowly varying waves inducing a nearly adiabatic electron motion. As a consequence, electrons orbits are deformed as the wave grows so that electrons with the same initial velocity are all trapped nearly simultaneously, and phase mixing at the origin of the decrease of $\nu$ is very efficient. In the situation considered by O'Neil, electrons with the same initial velocity are not all trapped by the wave, depending on their initial position. Moreover, by the time the wave amplitude has reached its constant value, the electrons orbits are essentially unperturbed, so that it takes more time for phase mixing to be effective. Hence, $\nu$ is less efficiently reduced to 0 in the O'Neil situation than in ours, and we find $\nuÊ\approx 0$ whenever $\int \omega_B dt \agt 6$, instead of  $\omega_B t \agt 30$ as found by O'Neil. 

\subsection{Three dimensional (3-D) space variation}
\label{3D}
We now discuss how, and when, 3-D effects may change the results derived previously, in the limit of a nearly unperturbed transverse electron motion. In case of a laser driven plasma wave, and when the laser electric field is polarized along the $y$ direction, one easily finds from Newton equations,
\begin{eqnarray}
v_y&=&v_{0y}+O(eA/m), \\
v_z&=&v_{0z}+O[(eA/m)^2/c],
\end{eqnarray}
where $A$ is the amplitude of the laser vector potential, while $v_{0y}$ and $v_{0z}$ are the unperturbed transverse velocities. Hence, the transverse motion may be considered as unperturbed provided that $eA/m \ll v_{th}$. This condition is fulfilled, for example, for typical laser and plasma conditions met in inertial confinement fusion. 

Let us now consider electrons with the same transverse velocities. Their contribution to $\I$, which we denote by $I_{1D}(v_{0y},v_{0z})$, is derived from the formulas of Sections \ref{time} { }Êand \ref{space}, provided that all quantities such as $\int_0^t \omega_B dt$, or $\gamma$, be now calculated in the frame moving at velocity $\vec{v}=v_\phi \hat{x}+v_{0y} \hat{y}+v_{0z}\hat{z}$ with respect to the laboratory frame since, in this frame, the electrons have no transverse motion. In particular, $\int_0^t \omega_B du$ now is, $\int_0^t \omega_B[x-\int_u^t v_\phi(t')dt',y-v_{0y}(t-u),z-v_{0z}(t-u),u]du$, and clearly assumes lower values than in 1-D. Indeed,  the electrons interact with the wave during a smaller time since, due to their transverse motion, they escape more rapidly from the region where the wave amplitude is significant. We therefore expect $\I$ to remain close to its linear value, and $\nu$ close to $\nu_L$, up to longer times in 3-D than in 1-D. Now, in order to calculate $\I$, we just need to sum over all contributions $I_{1D}(v_{0y},v_{0z})$, that is,
\begin{equation}
\I=\int_{-\infty}^{+\infty} I_{1D}(v_{0y},v_{0z}) f_0(v_{0y},v_{0z})dv_{0y}dv_{0z},
\end{equation}
where $f_0(v_{0y},v_{0z})$ is the unperturbed transverse distribution function. $\I$ assumes values significantly different from those derived in 1-D if $\vert \kappa_{y,z} v_{th} \vert \agt \vert \Gamma +\kappa v_{th} \vert$, where $\kappa_{y,z}Ê\equiv E_p^{-1} \partial_{y,z}E_p$, that is when the field amplitude variations experienced by the electrons is mainly due to the $y$ or $z$ dependence of $E_p$. Then, not only would $\nu$ decrease later as a function of time, but also more smoothly because the Heavyside-like function found in Sections \ref{exp} { }Êand \ref{space} is now convoluated with $f_0$. Hence, $\nu$ becomes a complicated operator of the transverse gradients of the wave amplitude, and may only be seen again as a damping rate if these gradients may be viewed as given parameters. For example, in case of a laser-driven plasma wave, the transverse dependence of $E_p$ is directly related to that of the laser intensity, due to its focusing inside of the plasma, and may therefore be considered as given. 

\section{Nonlinear frequency shift of a driven plasma wave}
\label{omega}
\begin{figure}
\centerline{\includegraphics[width=10cm]{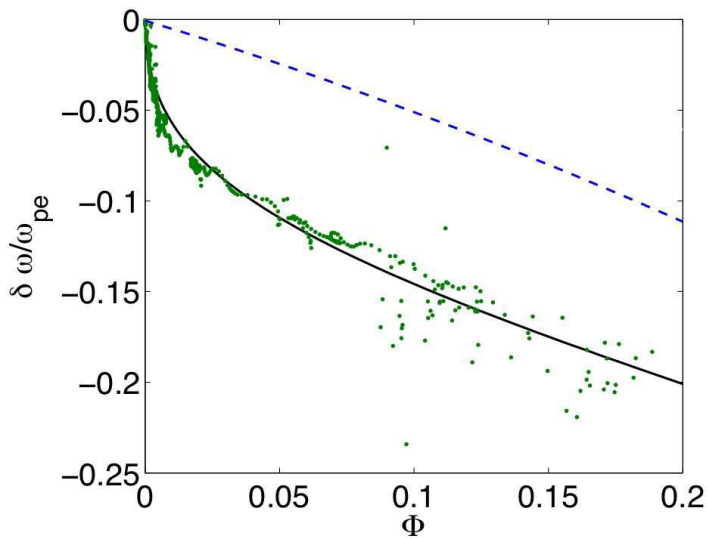}}
 \caption{\label{f9} The nonlinear frequency shift of the plasma wave, $\delta \omega$, as calculated numerically from Vlasov simulations of SRS (green dots), theoretically by solving Eq. (\ref{eq301}) (black solid line), and by solving Eq. (\ref{eq301}) with $\ad=1$ (blue dashed line), when $\kld \approx 0.52$. }
\end{figure}

In this Section, we briefly recall results discribed in Ref. \cite{benisti08} regarding the nonlinear frequency of a driven plasma wave. Plugging the definition (\ref{eq6}) of $\chi$ into Gauss law, one finds the following dispersion relation, 
\begin{equation}
\label{eq301}
1+\alpha_d\R=0,
\end{equation}
where
\begin{equation}
\label{eq302}
\alpha_d=\frac{1+2(E_d/E_p)\sin(\delta \varphi)+(E_d/E_p)^2}{1+(E_d/E_p)\sin(\delta \varphi)}.
\end{equation}
When the plasma wave is not driven, and $E_d=0$, $\ad=1$ and one recovers the usual dispersion relation $1+\R=0$. The linear value, $\alpha_{\text{lin}}$, of $\ad$ is chosen so as to correspond to the linearly most unstable wave against SRS, and its value results from the optimizing of two opposite trends. On one hand, it seems clear that it is easier to drive an electrostatic wave if this wave is a natural plasma mode.  Hence, $\alpha_{\text{lin}}$ should be close to unity. On the other hand, a wave grows more effectively if its Landau damping rate is small, that is if its phase velocity is large compared to the thermal one. Since, for a given wave number, $k$, the frequency $\omega$ derived from Eq. (\ref{eq301}) increases with $\ad$, we conclude that $\alpha_{\text{lin}}\agt 1$. Moreover, because the Landau damping rate increases with $\kld$, so does $\alpha_{\text{lin}}$. Now, from Eq. (\ref{eq1}) it is clear that, due to the decrease of $\I$ shown in the previous Sections, $E_d/E_p$ decreases as the plasma wave grows, which entails a rapid drop towards unity of $\ad$ and hence a rapid initial decrease of $\omega$. As a consequence, the frequency shift, $\delta \omega \equiv \omega-\omega_{\text{lin}}$, where $\omega_{\text{lin}}$ is the EPW linear frequency, is larger in magnitude than could be found by assuming that the wave was freely propagating \textit{i.e.}, by solving Eq. (\ref{eq301}) with $\ad=1$. This is illustrated in Fig. \ref{f9} which clearly shows that the initial drop in $\delta \omega$ is missed if one assumes $\ad=1$ when solving Eq. (\ref{eq301}). How to accurately calculate the nonlinear values of $\ad$ is explained in Ref. \cite{benisti08} and, accounting for the decrease of $\ad$ allowed us to derive values of $\delta \omega$ in very good agreement with those derived from Vlasov simulations of SRS, as shown in Fig. \ref{f9} when $\kld \approx 0.52$.

After the initial drop in $\omega$ due to that of $\ad$, the plasma wave frequency keeps on decreasing due to the nonlinear change in $\R$, which is calculated by making use of the adiabatic approximation. Then, the value we find for $\R$ in the limit of a vanishing wave amplitude is the same as that published, for example, in Refs. \cite{holloway,dewar,krapchev,rose}. However, unlike in these papers, we do find solutions to the dispersion relation when $\kld>0.53$, and for an infinitely small wave amplitude, because we solve Eq. (\ref{eq301}) with $\ad \neq 1$. Physically this means that, by sending a laser into a plasma it is always possible to drive an electrostatic wave, even with $\kld>0.53$ and slowly enough for an adiabatic estimate of $\R$ to be valid,  as shown in Ref. \cite{benisti08}. In order to calculate the nonlinear values of $\R$, by making use of the adiabatic approximation, we account for the nonlinear change of the phase velocity, which allows us to find solutions to the dispersion relation Eq. (\ref{eq301}) up to much larger values than if we had assumed that the wave frame was inertial, as was done in Refs. \cite{dewar,rose}.

\section{Application to stimulated Raman scattering}
\label{srs}
In this Section, we briefly discuss how our theoretical model applies to the studying of stimulated Raman scattering in the nonlinear regime, and we actually focus on the threshold of the so-called ``kinetic inflation''. This term was used in Ref. \cite{montgomerry} to design the regime where SRS reflectivity was experimentally found to be much larger than could be inferred from linear theory, a result which was attributed to the nonlinear reduction of the Landau damping rate. 

In its simplest version, SRS is a three wave process, an incident laser generating an electron plasma wave and a scattered electromagnetic wave. We assume that each of the electric field of these waves writes in terms of a slowly varying amplitude and an eikonal \textit{i.e.}, that the total electric field is,
\begin{equation}
\label{eq401}
\vec{E}_{tot}=E_p \sin(\varphi_p)\hat{x}+\hat{y}\left[E_l \sin(\varphi_l)+E_s \cos(\varphi_s)\right],
\end{equation}
where $E_p$, $E_l$ and $E_s$ are, respectively, the plasma, laser and scattered wave amplitude. We moreover require $\vert E_{p,l,s}^{-1}\partial_{t}ÊE_{p,l,s} \vert \ll \vert \partial_t \varphi_{p,l,s} \vert$ and  $\vert E_{p,l,s}^{-1}\partial_{x}ÊE_{p,l,s} \vert \ll \vert \partial_x \varphi_{p,l,s} \vert$. Then, in order to address the issue of SRS, one actually needs to solve three coupled envelope equations, one for each wave. It is actually more convenient to write these equations on complex quantities, which lets us define,
\begin{eqnarray}
E_p & \equiv & 2E_{0p}, \\
E_l & \equiv & 2E_{0l}e^{i(\kll x-\wll t)} e^{-i\varphi_l}, \\
E_s& \equiv &2E_{0s}e^{i(\ksl x-\wsl t)} e^{-i \varphi_s} e^{i\int_0^t\delta \omega(x,u) du}, 
\end{eqnarray}
\begin{figure}
\centerline{\includegraphics[width=14cm]{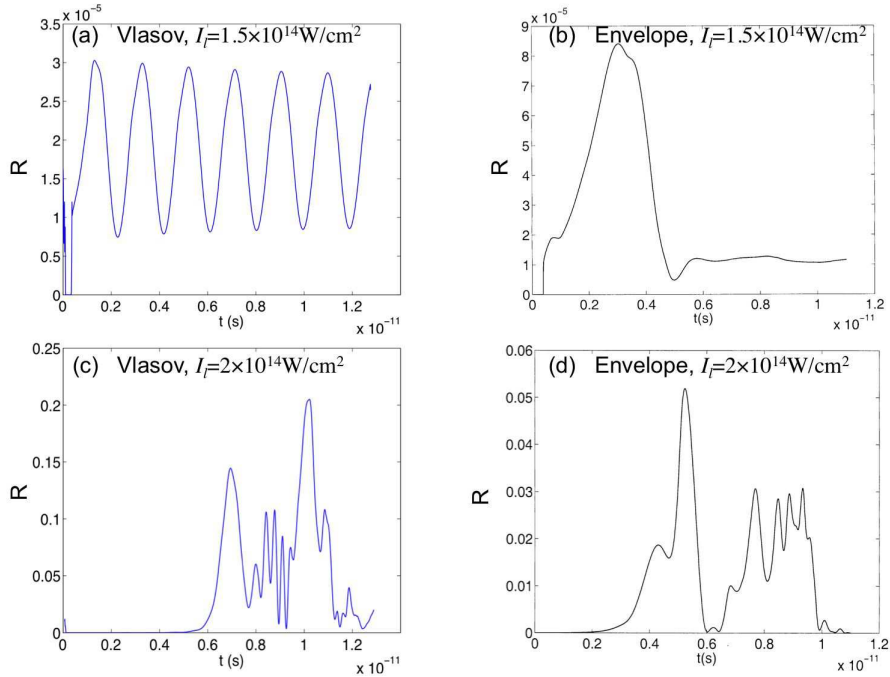}}
 \caption{\label{f10} Reflectivity, $R$, as a function of time when the laser intensity is $I_l=1.5 \times 10^{14}$W/cm$^2$, panel (a), as calculated using the Vlasov code ELVIS and, panel (b), using the envelope code BRAMA, and when the laser intensity is $I_l=2\times10^{14}$W/cm$^2$, panel (c), as given by a Valsov simulation and, panel (d), as given by our envelope code. }
\end{figure}
where $\kll$ and $\ksl$ are the linear values of the laser and scattered wave numbers, $k_{l,s}\equiv \partial_x \varphi_{l,s}$, $\wll$ and $\wsl$ are the linear values of the laser and scattered frequencies,  $\omega_{l,s}\equiv -\partial_t \varphi_{l,s}$, and $\delta \omega$ is the nonlinear frequency shift of the plasma wave, defined in Section \ref{omega}. Using Maxwell equations, and writing Gauss law as described in the previous Sections, we find the following equations, valid for a uniform plasma and in 1-D,  
\begin{eqnarray}
\label{eq402}
\frac{\partial E_{0p}}{\partial t}+v_{gp}\frac{\partial E_{0p}}{\partial x}+\nu E_{0p}&=&
\frac{\text{Re}(\Gamma_pE_{0l}E_{0s}^*)}{\dRv},\\
\label{eq403}
\frac{\partial E_{0s}}{\partial t}+v_{gs}\frac{\partial E_{0s}}{\partial x}+i\left [\delta \omega-v_{gs}\delta k\right]E_{0s}&=&\Gamma_s E_{0l}  E_{0p}^* , \\
\label{eq404}
\frac{\partial E_{0l}}{\partial t}+v_{gl}\frac{\partial E_{0l}}{\partial x} &=&-\Gamma_l E_{0s} E_{0p} ,
\end{eqnarray}
where, in Eq. (\ref{eq403}), $\delta k$ is the nonlinear wave number shift of the plasma wave, related to $\delta \omega$ by  the equation, $\partial_t \delta k = -\partial_x \delta \omega$, $v_{gl}$ and $v_{gs}$ are the usual group velocities of electromagnetic waves, $v_{gs}\equiv k_{l}c/\omega_{l}$, $v_{gs}\equiv k_{s}c/\omega_{s}$, as for $\Gamma_p$, $\Gamma_l$ and $\Gamma_s$, these are plain constants, $\Gamma_p=ek/m\omega_l\omega_s$, $\Gamma_s=ek/2m\omega_l$, and $\Gamma_l=ek/2m\omega_s$, where $k\equiv \partial_x \varphi_p$ is the plasma wave number. The envelope equations (\ref{eq402}-\ref{eq404}) are solved using the code BRAMA, which will be detailed in a forthcoming paper, and the results are compared to those of the Vlasov code ELVIS, Ref. \cite{strozzi}. In our simulations, either with the Vlasov or the envelope code, SRS results from the optical mixing of a laser, and a counterpropagating seed, as explained in Section \ref{space}. The ratio between the seed intensity $I_s(L)$ at the right end of the simulations box, and the laser intensity at the left end of the box, $I_L(0)$, is chosen to be $10^{-5}$. Figure \ref{f10} plots the reflectivity $R \equiv I_s(0)/I_L(0)$ as a function of time, calculated for a 1-D uniform plasma with electron temperature, $T_e=2$keV, electron density $n=0.1n_c$, and whose length is 100$\mu$m. The laser wavelength is 0.35 $\mu$m while the seed wavelength is 0.55 $\mu$m. When the laser intensity is $I_l=1.5\times10^{14}$W/cm$^2$, a linear theory would predict the reflectivity to be $R_{\text{lin}}\approx 2\times 10^{-5}$, and both the Vlasov and envelope codes find $R$ of the order of $10^{-5}$. By contrast, when $I_l=2\times10^{14}$W/cm$^2$, while the linear value of the reflectivity is $R_{\text{lin}} \approx 3\times10^{-5}$, the reflectivity calculated either with the Vlasov or the envelope code is of the order of 10\%, as can be seen in Fig. \ref{f10}. This Figure also shows some discrepancies in the actual values of the reflectivity predicted by the two different codes, whose origin will be discussed in a future paper and is way beyond the scope of this article. However, as regards the threshold for inflation, both codes agree that the threshold intensity lies between $1.5\times10^{14}$W/cm$^2$ and $2\times10^{14}$W/cm$^2$, while the envelope code is about 5000 faster in providing this result. Hence, using the theoretical model described in the previous Sections, we built a powerful tool to predict when stimulated Raman scattering is negligible, which an important issue for inertial confinement fusion (see for example Ref. \cite{cavailler}).

\section {Conclusion}
In this paper, we investigated how efficiently an electron plasma wave (EPW) could be externally driven. This led us define the nonlinear group velocity, $v_g$, and Landau damping rate, $\nu$, of a driven plasma wave, which are terms appearing naturally in the envelope equation for the wave amplitude. We provided a practical analytic formula for $\nu$, and found the unexpected result that $\nu$ assumed nearly constant values before abruptly dropping to zero, and that this drop in $\nu$ occurred simultaneously with a rapid increase of $v_g$ towards the wave phase velocity, and a decrease of the coupling constant between the plasma wave and the driving field. We moreover unambiguously showed, without resorting to complex contour deformation, that a plasma wave, first driven by laser at a small enough amplitude and then freely propagating, would damp at the rate predicted by Landau. This then imposes restrictions for non-Landau damping, as predicted by Belmont \textit{et al.} in Ref. \cite{belmont}, to indeed occur in actual experiments. All these results stem from our theoretical derivation of $\I$, which directly follows from the investigation of the nonlinear electron motion. The expression found  for $\I$ actually results from the matching of two very different estimates, a perturbative one for small amplitudes, and one relying on the adiabatic approximation and valid whenever $\nu \approx 0$. This yields values for $\I$ in excellent agreement with those either inferred from test particle simulations or from Vlasov simulations of stimulated Raman scattering (SRS).

We moreover discussed in this article the nonlinear frequency shift, $\delta \omega$, of a driven plasma wave and found that $\vert \delta \omega \vert$ was much larger than could be derived by assuming that the wave was freely propagating. We moreover showed that no physical effect could be attributed to the increase of $\kld$ above 0.53, unlike what could be inferred from Ref. \cite{holloway}. This emphasizes the importance of specifying the way a plasma wave has actually been generated in order to discuss its nonlinear properties.

Our results regarding both, the  EPW envelope equation and its nonlinear frequency shift, allow us to study SRS in the nonlinear regime. In particular, we investigated the threshold of the so-called kinetic inflation, a regime where the SRS reflectivity is much larger than predicted by linear theory. This threshold is a very important parameter for inertial confinement fusion because, below it, one is assured that SRS reflectivity would be very low and therefore that SRS would not affect the fusion efficiency. Using  our model when the plasma is homogeneous, and in a 1-D geometry, we found values for the  inflation threshold in very good agreement  with those derived from Valsov simulations, but within a much smaller computing time. This shows the potentiality of our model to address more complicated physics situations.

In conclusion, we derived very precisely the nonlinear properties of a driven electron plasma wave, which allowed us to discuss the generality of previous results on this topic, which is a long standing, and basic issue in plasma physics. We moreover applied our results to the studying of stimulated Raman scattering, and to the threshold for kinetic inflation, which is an important issue for inertial confinement fusion.

\label{conclusion}
\appendix
\section{Hamiltonian perturbative analysis}
\label{A}
\setcounter{equation}{0}
\newcounter{app}
\setcounter{app}{1}
In this Appendix, we use a first order Hamiltonian perturbative analysis to approximate the motion of an electron acted upon by a longitudinal wave whose electric field is $E \equiv E_0(t)e^{i \varphi(x,t)}+c.c.$, and whose frequency, $\omega$, and wave number, $k$, are defined by $k = \partial_x \varphi$, $\omega=-\partial_t \varphi$. In the dimensionless variables, $\tau = t/kv_{th}$, $\varphi(\tau) = \varphi[x(\tau),\tau]$ 	and $v = v_{th}^{-1}dx/dt$, where $v_{th}=\sqrt{T_e/m}$ is the thermal velocity, the electron dynamics derives from the Hamiltonian,
\begin{equation}
H=\frac{(v-v_\phi)^2}{2}+(i\Phi e^{i\varphi}+c.c.)
\end{equation}
 where $\Phi=eE_0/kT_e$, and $v_\phi = \omega /kv_{th}$. The perturbative calculation consists in defining a canonical change of variables $(\varphi,v) \rightarrow (\varphi',v')$ such that $v'$ is a constant of motion, at least at first order in the wave amplitude. The change of coordinates is defined using a generative function, $F(\varphi,v')$, and is
\begin{eqnarray}
\label{x}
\varphi'&=&\varphi+\partial_{v'} F, \\
\label{v}
v&=&v'+\partial_\varphi F.
\end{eqnarray}
Then, $\varphi \approx \varphi_0+(v_0-v_\phi)\tau +\delta \varphi$, where $\varphi_0$ and $v_0$ are constant, and
\begin{equation}
\delta \varphi=-\partial_v' F.
\end{equation}
In the new variables, the new Hamiltonian is,
\begin{equation}
H'=H+\frac{\partial F}{\partial t}= \frac{(v'+\partial_\varphi F-v_\phi)^2}{2}+(i\Phi e^{i\varphi}+c.c.)+\frac{\partial F}{\partial t}.
\end{equation}
The generative function, $F$, is then chosen so as to cancel the term $i\Phi e^{i\varphi}+c.c.$, so that, at first order in $\Phi$, it needs to solve, 
\begin{equation}
\label{F_1}
(v'-v_\phi)\frac{\partial F}{\partial \varphi}+\frac{\partial F}{\partial t}=-i\Phi e^{i\varphi}+c.c.
\end{equation}
We now assume that, at $\tau=0$, the wave amplitude is infinitesimal, so that $\delta \varphi=F=0$. Then, the solution of Eq. (\ref{F_1}) is,
\begin{equation}
F=-ie^{i\varphi}\int_0^\tau \Phi(u)e^{iw(u-\tau)}.c.c.,
\end{equation}
where we have denoted $w=v'-v_\phi$. Then,
\begin{eqnarray}
\nonumber
\delta \varphi&=&-ie^{i\varphi}\partial_w \left(\int_0^\tau \Phi(u)e^{iw(u-\tau)}.c.c.\right) \\
&\approx&-ie^{i(\varphi_0+w \tau)}\partial_w \left(\int_0^\tau \Phi(u)e^{iw(u-\tau)}.c.c.\right) 
\end{eqnarray}

\section{Approximate expression for $\Ic$.}
\label{B}
\setcounter{equation}{0}
\setcounter{app}{2}

In this Appendix, we give an approximate expression of 
\begin{equation}
\label{b1}
\chi_a=\frac{if'_0(v_\phi)}{(\kld)^2\Phi(\tau)}\int_0^\tau \Phi(u)(u-\tau) \int_{\vert w \vert >\Vl} iw e^{iw(\xi-\tau)}dw du,
\end{equation}
in the limit $\Vl \ll \tau_\Phi^{-1}$, where $\tau_\Phi$ is the typical timescale of variation of $\Phi$. From the results of Section \ref{time}, it is clear that $\Ic=-(\kld)^{-2}f'_0(v_\phi)[\pi+\delta \chi_a]$, with,
\begin{equation}
\label{b2}
\delta \chi_a= \Phi(\tau)^{-1}\int_0^\tau (u-\tau)\Phi(u) \partial_u G(u-\tau)du
\end{equation}
where
\begin{equation}
\label{b3}
G(u-\tau) = \int_{-\Vl}^{\Vl}e^{iw(u-\tau)}dw.
\end{equation}
Clearly, the timescale of variation of $G$ is $\Vl^{-1}$, while $\partial_uG \vert_{u=\tau}=0$, and $\partial^2_{u^2}G \vert_{u=\tau}=2\Vl^3/3$. Then, integrating (\ref{b2}) three times by parts yields
\begin{eqnarray}
\nonumber
\delta \chi_a&=&-\frac{4\Vl^3}{3\Phi(\tau)}\int_0^\tau \int_0^u\int_0^\xi \Phi(\xi')d\xi'd\xi du \\
\label{b6}
&&+\Phi(\tau)^{-1}\int_0^\tau \left[(u-\tau)\partial^4_{u^4}G+3\partial^3_{u^3}G \right]Ê\left( \int_0^u \int_0^\xi \int_0^{\xi'} \Phi(\xi'') d\xi'' d\xi' d\xi\right) du.
\end{eqnarray}
Clearly, the last term in the right-hand side of Eq. (\ref{b6}) is of the order $(\Vl\tau_\Phi)$ times the first one, and is therefore negligible in the limit $\Vl \ll \tau_\Phi^{-1}$.

\end {document}